\documentclass[12pt]{iopart}
\usepackage{iopams}
\usepackage{graphicx}
\usepackage{bm}
\usepackage{url}
\begin{document}

\title[Spectral super-resolution]{Spectral super-resolution in
  metamaterial composites}

\author{J. Helsing$^1$, R. C. McPhedran$^2$ and G. W. Milton$^3$}
\address{$^1$ Numerical Analysis, Centre for Mathematical Sciences,
  Lund University, Box 118, SE-221 00 Lund, Sweden} \address{$^2$
  CUDOS, School of Physics, University of Sydney, NSW 2006, Australia}
\address{$^3$ Department of Mathematics, University of Utah, Salt Lake
  City, UT 84112, USA}
\begin{abstract}
  We investigate the optical properties of periodic composites containing
  metamaterial inclusions in a normal material matrix. We consider the
  case where these inclusions have sharp corners, and following
  Hetherington and Thorpe, use analytic results to argue that it is then possible
  to deduce the shape of the corner (its included angle) by 
  measurements of the absorptance of such composites when the scale size of the 
  inclusions and period cell 
  is much finer than the wavelength. These analytic arguments are supported
  by highly accurate numerical results for
  the effective permittivity function of such composites as a
  function of the permittivity ratio of inclusions to matrix. The results
  show that this function has a continuous spectral component with
  limits independent of the area fraction of inclusions, and with the same limits for both
  square and staggered square arrays.
\end{abstract}

\submitto{\NJP}
\maketitle

\section{Introduction}

This paper links themes evoked in two classic papers, one in
mathematics \cite{kac} and the other in physics \cite{pendry2000}. The
first of these poses the question as to whether the spectral
content of the radiation from a body can reveal its shape. The second shows
that the use of negatively-refracting metamaterials in a plane slab
can lead to a super-resolving perfect lens, also known as a superlens. We will consider a
two-dimensional composite material, composed of polygonal inclusions
made of a metamaterial (by which we mean an artificial material with a
dielectric contant which has a negative real part and a very small imaginary part)
and placed in a positive dielectric matrix material. We
will show that, in the spirit of Pendry, the metamaterial makes
possible resolution of an important structural feature of the
inclusions, irrespective of how much smaller than the wavelength they
are. We will also show that, in the spirit of Kac, this feature
relates to the shape of the inclusion, being in fact the corner angle
of the polygon, and that it is deduced from spectral measurements on
the composite. The fact that a spectral feature could be determined by
corner shape, independent of (say) the area fraction of inclusions,
was first suggested by Hetherington and Thorpe~\cite{handth}, on the
basis of an elegant argument and numerical evidence for dilute
composites.

We will base our demonstration firstly on analytic results relating to
the spectrum of the effective dielectric permittivity function
$\epsilon_{\rm eff}$ of the composite material, and secondly on
remarkably accurate numerical results for this spectrum obtained using
a new technique. Note here that we are using the word "spectrum" in
two related, but slightly different senses. In the previous paragraph,
its usage meant that the absorption of electromagnetic waves by the
composite was being determined as a function of wavelength, ranging
over an appropriately-wide band. In the first sentence of this
paragraph we referred to the distribution of singularities of the
function $\epsilon_{\rm eff}$, giving the effective permittivity of a
composite having a specified geometry as a function of the ratio
$\sigma=\epsilon_1/\epsilon_2$ of the permittivities of inclusions and
matrix. The relation between these usages is that, as wavelength
varies so does the ratio $\sigma$, so that measurements of (say)
optical absorption by a composite over a suitable wavelength interval
can reveal details of the singularities of the function $\epsilon_{\rm
  eff}$.

The numerical results for the singularity spectrum of $\epsilon_{\rm
  eff}$ reveal that it has a continuous part which runs between upper
and lower limits of $\sigma$ which do not vary at all with the area
fraction of the inclusions. It is complemented by a discrete spectrum
of poles which does evolve with area fraction. This evolution is in
fact necessary, since the continuous spectrum for touching square
inclusions in a checkerboard arrangement occupies the entire negative
real axis of $\sigma$, but for non-touching square inclusions is
confined to the interval $-3\le\sigma\le-1/3$. The animations we give
show how this transformation is achieved: the discrete spectrum
becomes more and more dense as the touching configuration is
approached, to supply the required spectral extension, as anticipated
by one of the authors~\cite{miltbook}.

The results we give here are interesting in the insights they give
into the connection between metamaterials and super-resolution. They
are also important in furthering our understanding of the connection
between inclusion shape, geometrical arrangement and spectral
properties of the effective permittivity function. This connection
helps in the design of structures having enhanced absorption over a
wide wavelength range for applications in photothermal or photovoltaic
captors \cite{mcpandp80,pendry2010a,pendry2010b,pendry2010c}, or
offering strongly enhanced local fields for applications like sensing
or nonlinear optical elements.

We give a brief overview in Section~\ref{sec:over} of some of the
important properties of the function $\epsilon_{\rm eff}(\sigma)$,
including analytic results relating to the continuous part of the
spectrum and some numerical investigations of both the discrete and
continuous parts of the spectrum. In Section~\ref{sec:numer}, we
describe the method which enables accurate calculation of the spectrum
of $\epsilon_{\rm eff}$, and give numerical results illustrating its
convergence, even on the negative real axis of $\sigma$. In
section~\ref{sec:anim}, we discuss the animations which are given in
the Supplementary Material to this paper, and the physical
consequences of the behaviour they show. We give a discussion and
concluding remarks in Section~\ref{sec:disc}.

\section{Properties of the effective permittivity function for composites}
\label{sec:over}

We will now give a concise review of what is known about the
properties of the effective dielectric permittivity function
$\epsilon_{\rm eff}$, for composites made of two materials with
dielectric permittivities $\epsilon_1$ and $\epsilon_2$, with the
former corresponding to a disconnected inclusion phase and the latter
to a continuous matrix phase. We assume the geometry has cubic symmetry (in three dimensions)
or square symmetry (in two dimensions) so that $\epsilon_{\rm eff}$ is scalar valued, i.e. the 
effective dielectric tensor equals $\epsilon_{\rm eff}I$.
This review builds on that given in
Perrins and McPhedran \cite{perrmcp2010}.

The calculation of $\epsilon_{\rm eff}$ for a given geometry is homogeneous of
degree 1 in the variables $\epsilon_1$ and $\epsilon_2$, so we can rescale to make $\epsilon_{\rm eff}$
a function of a single complex variable, the permittivity ratio
$\sigma=\epsilon_1/\epsilon_2$:
\begin{equation}
\epsilon_{\rm eff}(\epsilon_1,\epsilon_2)=
\epsilon_2\epsilon_{\rm eff}(\sigma,1)=
\epsilon_2\epsilon_{\rm eff}(\sigma)\,.
\label{eq1}
\end{equation}
From now on we let $\epsilon_{\rm eff}$ be an abbreviated notation for
the effective dielectric permittivity 
function $\epsilon_{\rm eff}(\epsilon_1,\epsilon_2)$ and we let $\epsilon_{\rm eff}(\sigma)$
denote its scaled counterpart, the effective relative dielectric
permittivity function, as defined via~(\ref{eq1}).

To determine $\epsilon_{\rm eff}(\sigma)$ for a given geometry, we
need to solve an electrostatic transport problem repeatedly for
various $\sigma$. More precisely, we need to solve Laplace's equation
for the potential $V$
on a periodic domain with a periodic electric field $-\nabla V(x,y)$
having a prescribed average value, and boundary
conditions of continuity of $V$ and its normal
flux $\epsilon \partial V\partial n$ at interfaces between materials.
The theory of the function $\epsilon_{\rm eff}(\sigma)$ becomes
particularly elegant when we deal with two-dimensional problems, in
which $V=V(x,y)$ becomes a function in the plane. We may then apply
the apparatus of complex-variable theory to the calculation of $V$,
and thus to $\epsilon_{\rm eff}(\sigma)$. For the case of a
doubly-periodic array of inclusions ${\cal C}$ with unit cell $\cal
U$, and square symmetry, the effective permittivity may be defined as~\cite{berg78}
\begin{equation}
\epsilon_{\rm eff}=\frac{\int_{\cal U}\epsilon|\nabla V(x,y)|^2 dx dy}
{|\int_{\cal U}\nabla V(x,y) dx dy|^2}\,,
\label{epsdef}
\end{equation}
where the integral in the numerator includes contributions ${\cal E}_1$ from the
inclusion region and ${\cal E}_2$ from the matrix region. Except for
occasional comments on effective permittivity in three-dimensions, we
will concentrate on two dimensions, which corresponds to arrays of
cylinders of arbitrary cross-section, with the average field aligned
in the $(x,y)$ plane. The area fractions of the two phases will be
denoted $p_1$ and $p_2$.

Since the geometry has square symmetry, Keller's
Theorem~\cite{keller} gives
\begin{equation}
\epsilon_{\rm eff}(\sigma)\epsilon_{\rm eff}(1/\sigma)=1\,.
\label{eq2}
\end{equation}
This equation then pairs zeros of $\epsilon_{\rm eff}(\sigma)$ at
values $\sigma_0$ with poles at values $\sigma_p=1/\sigma_0$. Of
course, from~(\ref{eq2}), zeros of $\epsilon_{\rm eff}(\sigma)$
require that the contributions ${\cal E}_1$ and ${\cal E}_2$ add up to
zero. Since, with $\epsilon_2=1$, ${\cal E}_2$ is real and positive, this means ${\cal E}_1$
must be real and negative, and so $\epsilon_2=\sigma$ must be real and negative
at any zero of $\epsilon_{\rm eff}(\sigma)$, and thus at any pole as
well. Bergman~\cite{berg78} proved this property for both
two-dimensional and three-dimensional composites. Bergman also
recognized that even though the transport problem when the average electric field $-\nabla V(x,y)$
is prescribed does not have a solution at a pole, it should have a solution at a pole when instead
the average displacement field $-\epsilon\nabla V(x,y)$ is prescribed.
Milton~\cite{milt79} proved
that for the value $\sigma=-1$ the electrostatic problem of an array
of circular cylinders, with either a prescribed value of the average electric field
$-\nabla V(x,y)$ or a prescribed average value of the displacement field $-\epsilon\nabla V(x,y)$,
 does not have a solution, compare with the
discussion of~(\ref{eq:int1}) below. This suggests that $\sigma=-1$ is
either an essential singularity or lies on a branch-cut of the
function $\epsilon_{\rm eff}(\sigma)$ for arrays of circular cylinders.

The fact that branch-cuts cannot be in the upper or lower half-planes,
but must lie exactly on the negative real axis of $\sigma$ was proved
by one of the authors \cite{milt1981a}, using the relationship between
composite materials and resistor networks. A rigorous justification of
the spectral representation for $\epsilon_{\rm eff}(\sigma)$ was given
by Golden and Papanicolaou~\cite{goldpapa}. In general the function $\epsilon_{\rm eff}(\sigma)$
has the representation
\begin{equation}
\epsilon_{\rm eff}(\sigma)=a_0+a_1\sigma+\int_{-\infty}^0\frac{d\mu(\tau)}{\tau-\sigma},
\label{spect}
\end{equation}
where $a_1$ and the spectral measure $d\mu(\tau)$ are non-negative. The support of
$d\mu(\tau)$ is the spectrum. The spectral measure can be recovered from the 
values that the imaginary part of $\epsilon_{\rm eff}(\sigma)$ takes near the negative real $\sigma$-axis since the
integral of any smooth test function $g(\tau)$ with respect to the measure $d\mu(\tau)$
is given by
\begin{equation}
\int_{-\infty}^0g(\tau)d\mu(\tau)=\lim_{\matrix{\delta\to 0 \cr\delta> 0}}
\frac{1}{\pi}\int_{-\infty}^0g(\tau)\Im\epsilon_{\rm eff}(\tau+i\delta)~d\tau.
\label{measrec}
\end{equation}

The discrete spectrum of $\epsilon_{\rm eff}(\sigma)$ is readily
exhibited numerically. This has been done for arrays of spheres by
Bergman~\cite{berg79} and for arrays of circular cylinders by McPhedran and
McKenzie~\cite{mandm80}, with both studies showing that the poles and zeros
of $\epsilon_{\rm eff}(\sigma)$ converge to an essential singularity at
$\sigma=-1$.

We focus now on what can be said about the continuous spectrum of
$\epsilon_{\rm eff}(\sigma)$. One simple geometry for which a result
is immediately apparent is the checkerboard, for which
Dykhne~\cite{dykhne} obtained from Keller's theorem~(\ref{eq2}) the
exact result
\begin{equation}
\epsilon_{\rm eff}(\sigma)=\sqrt{\sigma}\,.
\label{dyk}
\end{equation}
This then exhibits a branch cut along the entire negative real axis of
$\sigma$.

An exact result can also be obtained for the polarizability $\alpha$
of a pair of touching cylinders~\cite{mcpandp80}. Using an inversion
of coordinates about the contact point, the touching cylinders may be
transformed into a slab of matrix material with permittivity
$\epsilon_2=1$ surrounded by two
half-planes filled with material with permittivity $\epsilon_1=\sigma$.
Introducing the parameter
\begin{equation}
\lambda=\frac{\sigma-1}{\sigma+1}\,,
\label{eq:lmb}
\end{equation}
it is easy to show using the method of images that the polarizability
for a pair of touching cylinders of radius $a$ for the case of the
applied field parallel to the line connecting cylinder centres is
\begin{equation}
\alpha=4\pi a^2\sum_{l=1}^\infty
\frac{\lambda^l}{l^2}\,.
\label{dilog1}
\end{equation}
We see that the series in (\ref{dilog1}) converges provided
$|\lambda|<1$, i.e. for real $\sigma$, $\sigma >0$. However, we can
obtain a meaningful result even when this is not the case by the
technique of analytic continuation, since the series in (\ref{dilog1})
is a known transcendental function, called the dilogarithm, and
denoted ${\rm Li_2}$. Thus, we can replace (\ref{dilog1}) by
\begin{equation}
\alpha=4\pi a^2{\rm Li}_2(\lambda)\,.
\label{dilog2}
\end{equation}
The properties of the dilogarithm function are that it has a branch
cut running from $\lambda=1$ to $\lambda=\infty$, across which the
discontinuity in the imaginary part of ${\rm Li_2}(\lambda)$ is
$2\pi\log[\Re\{\lambda\}]$. The branch cut in the plane of relative
permittivity runs from $\sigma=-\infty$ to $\sigma=-1$. If the
direction of the applied field is perpendicular to the line of
centres, the branch cut runs from $\sigma=-1$ to $\sigma=0$.

We next consider arrays of square inclusions, for which we have
already mentioned the Dykhne result (\ref{dyk}). A generalization of
this for an array in which the square unit cell was divided into four
equal squares with dielectric permittivities $\epsilon_1$, $\epsilon_2$,
$\epsilon_3$, and $\epsilon_4$, was conjectured by Mortola and
Steff\'{e}~\cite{mortstef85}:
\begin{equation}
\epsilon_{\rm eff}=\left[
\frac{(\epsilon_2+\epsilon_3)(\epsilon_4+\epsilon_1)}
     {(\epsilon_1+\epsilon_2)(\epsilon_3+\epsilon_4)}\right]^{1/2}
\left(\frac{\epsilon_1\epsilon_2\epsilon_3+\epsilon_1\epsilon_2\epsilon_4
      +\epsilon_1\epsilon_3\epsilon_4+\epsilon_2\epsilon_3\epsilon_4}
{\epsilon_1+\epsilon_2+\epsilon_3+\epsilon_4}\right)^{1/2}\,.
\label{eq3a}
\end{equation}
This conjecture was proved independently by Milton \cite{milt01} and
Craster and Obnosov \cite{crastob2001}.

\begin{figure}
\begin{center}
\includegraphics[width=12cm]{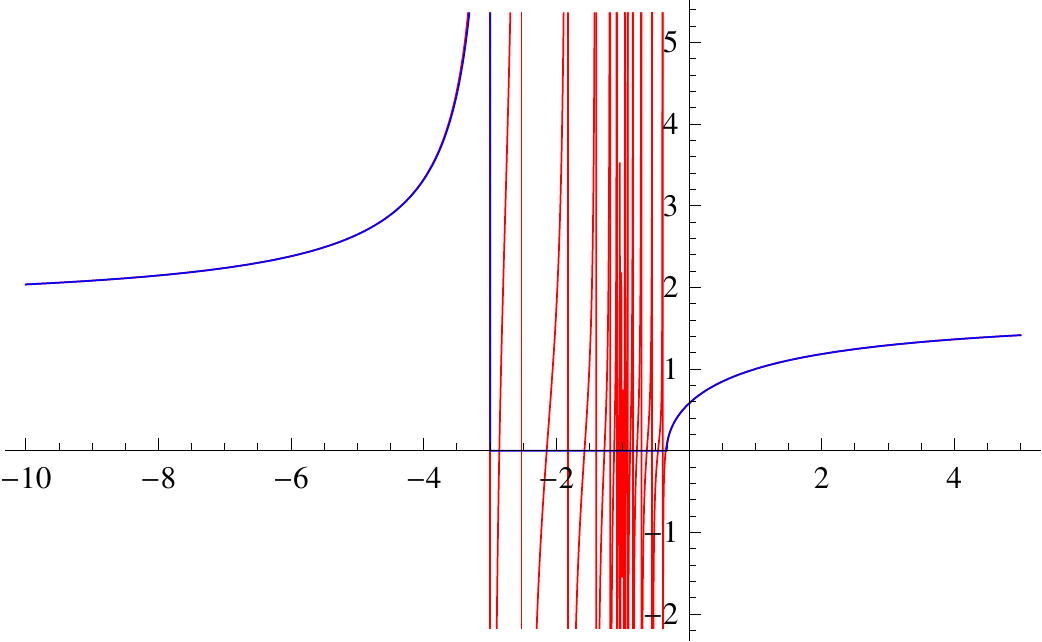}
\end{center}
\caption{The blue curve gives the real part of the formula (\ref{eq4})
  for the effective relative dielectric permittivity of the square
  array of square cylinders, while the red curve gives the result of a
  mode-matching method \cite{perrmcp2010}.}
\label{fig1}
\end{figure}

We will consider a particular sub-case of this result, due to Obnosov
\cite{obnos}: an array of square cylinders with area fractions
$p_1=0.25$ and $p_2=0.75$, for which (\ref{eq3a}) gives
\begin{equation}
\epsilon_{\rm eff}(\sigma)=\sqrt{\frac{1+3\sigma}{\sigma+3}}\,. 
\label{eq4}
\end{equation}
This formula yields a spectrum consisting solely of a branch-cut
running from $\sigma=-3$ to $\sigma=-1/3$. We compare the result given
by this formula for the real part of $\epsilon_{\rm eff}(\sigma)$ with
the result of a numerical mode matching procedure in Fig.~\ref{fig1}.
This comparison reveals the difficulty of evincing details of the
spectrum using numerical methods: the mode matching technique
approximates the branch cut by a discrete set of poles, which becomes
more dense as the number of modes increases. However, it is difficult
to distinguish between branch cuts and sets of poles concentrating
around an essential singularity by such methods. Furthermore, the mode
matching method failed to give clear indications of the spectrum for
area fractions of cylinders distinct from $p_1=0.25$.

The behaviour of fields near corners of inclusions in a matrix of
differing dielectric permittivity and magnetic permeability was
treated in electromagnetism by J. Meixner~\cite{meixner}. As well as
obtaining the exponents which characterize the singularity of the field
components near the edge, Meixner pointed
out that exactly the same formulae could be applied in electrostatics
and magnetostatics. Hetherington and Thorpe~\cite{handth} analysed the
behaviour of fields near corners in electrostatics and magnetostatics,
apparently without knowledge of Meixner's paper. They however pointed
out the link between field behaviour near corners and the nature of
the singularity spectrum in composites containing inclusions with
corners. Let us suppose that the electrostatic potential varies with
distance $r$ from a corner with an included angle $2\psi$ as
$r^\beta$, and that the permittivity ratio inside the inclusion to
that outside is $\sigma$. Then~\cite{meixner,handth} $\beta$ is found
by solving a transcendental equation:
\begin{equation}
\frac{\tan [\beta(\pi -\psi)]}{\tan (\beta \psi)}=-\sigma\,,
\label{transeq}
\end{equation}
or the same equation with $\sigma$ replaced by $1/\sigma$. For
$\sigma$ real, the solution $\beta$ of (\ref{transeq}) is either pure
real or pure imaginary.

To clarify the behaviour of fields, consider the case of $\psi=\pi/4$,
corresponding to a $90^\circ$ corner. Then (\ref{transeq}) gives
\begin{equation}
\cos(\beta\pi/2)=\frac{(\sigma -1)}{2(\sigma +1)}\,.
\label{squcorn}
\end{equation}
From this, we see that the right-hand side exceeds one in magnitude
for $\sigma$ lying between $-3$ and $-1/3$, and that $\beta$ is then
pure imaginary, corresponding to a solution for the potential which
oscillates with $r$, and the oscillations become ever more rapid as
$r$ tends to zero. The electric field is given by the spatial
derivative of the potential, and it diverges like $1/r$ multiplied by
the oscillating term as $r\rightarrow 0$. The branch cut region here
is that where $\beta$ is imaginary, i.e., between $\sigma=-3$ and
$\sigma=-1/3$. Hetherington and Thorpe \cite{handth} postulate that
for an $m$ sided regular polygon, there will be a branch cut running
between $\sigma=(2+m)/(2-m)$ and $\sigma=(2-m)/(2+m)$. They made this
assertion after recognizing that in this interval of $\sigma$ the 
surface charge (rather that just the surface charge density) near the
corner is infinite, which is unphysical.

Another argument for the position of this branch cut was put forward by
one of the authors \cite{miltbook}. 
In order that $\epsilon_{\rm eff}$ have a significant imaginary part
when $\epsilon$ has a very small imaginary part we see from (\ref{epsdef})
that the electric field $-\nabla V$ must be close to losing its square integrability.
From the asymptotic form of $V$ near the corner one finds that this happens when the
imaginary part of $\sigma$ is very small and
real part of $\sigma$ is between $(2+m)/(2-m)$ and $(2-m)/(2+m)$. To elucidate this further
for the case of a  $90^\circ$ corner we 
solve (\ref{squcorn}) with $\beta=\beta'+i\beta''$ and $\sigma=\sigma'+i\sigma''$ where 
$\beta',\beta'',\sigma'$ and $\sigma''$ are real, $-3<\sigma'<-1/3$, and
$\sigma''$ and $\beta'$ are both very small and positive. This gives 
\begin{equation}
\beta'\approx\frac{4\sigma''}{(\sigma'+1)\sqrt{(\sigma'+3)(3\sigma'+1)}}.
\label{limbetsig}
\end{equation}
Using polar coordinates $(r,\theta)$ near the corner, the potential $V$ 
scales as $r^{\beta}$ as $r\to 0$, and so $|\nabla V|^2 $ will be close to $r^{2\beta'-2}|g(\theta,\sigma')|^2$  
for some function $g(\theta,\sigma')$. It follows that with $\epsilon_1=\sigma$ and $\epsilon_2=1$
the imaginary part of ${\cal E}_1$ (which is a measure of the power dissipation in 
the composite) has a contribution near the corner from inside the radius $r=r_0$ of
\begin{eqnarray}
\int~d\theta\int_0^{r_0}\sigma''|\nabla V|^2r~dr & = & \frac{\sigma''r_0^{2\beta'}}{2\beta'}\int|g(\theta,\sigma')|^2~d\theta
\nonumber \\
& = & \frac{r_0^{2\beta'}}{8}|(\sigma'+1)\sqrt{(\sigma'+3)(3\sigma'+1)}|\int|g(\theta,\sigma')|^2~d\theta \nonumber \\ &~&
\label{engloss}
\end{eqnarray}
where the integral over $\theta$ is only over those angles in the inclusion. Thus, provided $g(\theta,\sigma')$ 
is non-zero, the contribution of the corner
to the imaginary part of ${\cal E}_1$ remains non-zero even in the limit $\sigma''\to 0$ (and goes to zero when $\sigma'<-3$ or $\sigma'>-1/3$ since then $\sigma''/\beta'\to 0$).

A corner is not the only geometric feature which can act as a significant energy absorber when 
the imaginary part of the dielectric constant goes to zero. The center of a sphere with a dielectric constant
$\lambda_1$ in the radial direction and dielectric constant $\lambda_2$ in the tangential direction acts as an absorber
when $\lambda_2/\lambda_1$ approaches real values less than $-1/8$: see figure 4 in the paper of 
Qui and Luk’yanchuk \cite{quiluk} (which shows that this energy absorbing feature extends beyond the quasistatic limit)
and see also the related discussion on page 239 of \cite{miltbook} (where there is an error as the limit $\delta\to -1/2$ should have been taken rather 
than the limit $\delta\to 0$). 

\section{Numerical method}
\label{sec:numer}
We now  describe a numerical method stable and
accurate enough to verify the conjecture of Hetherington and Thorpe
for the case $m=4$, and to show the spectral evolution as a function
of area fraction for composites with square inclusions.

\begin{figure}[t]
\begin{center}
  \includegraphics[width=35mm]{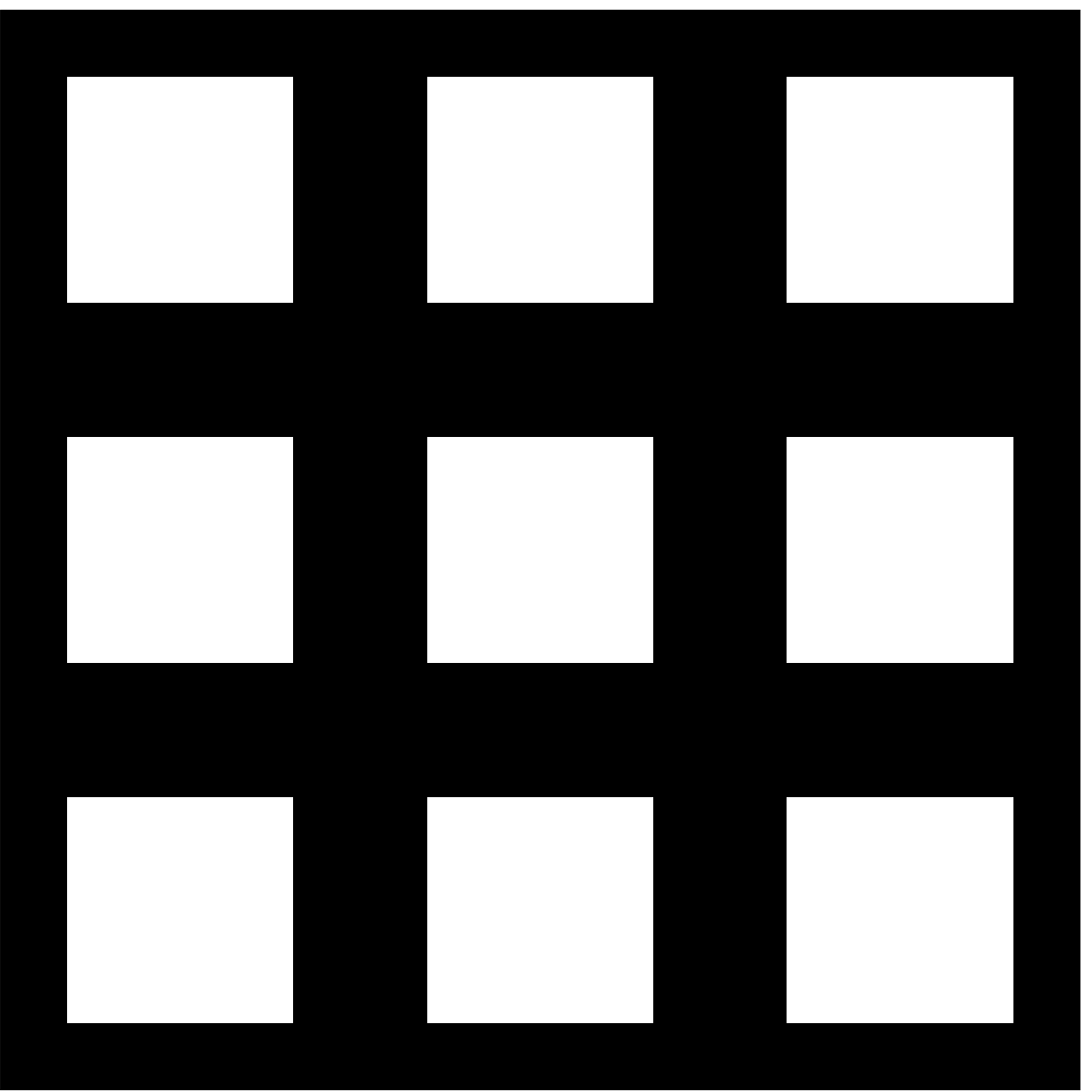}
  \includegraphics[width=35mm]{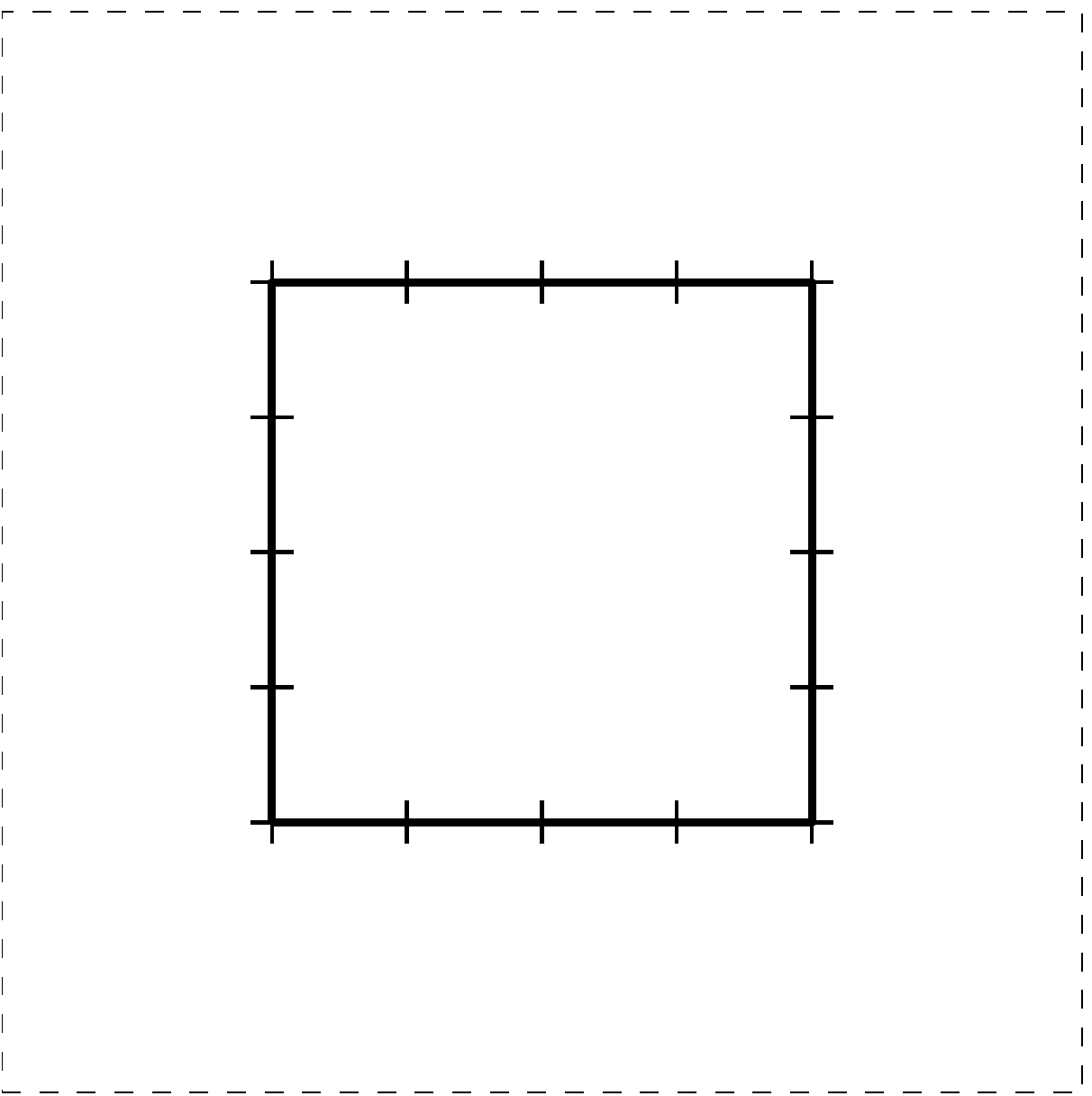}
  \hspace{2mm}
  \includegraphics[width=35mm]{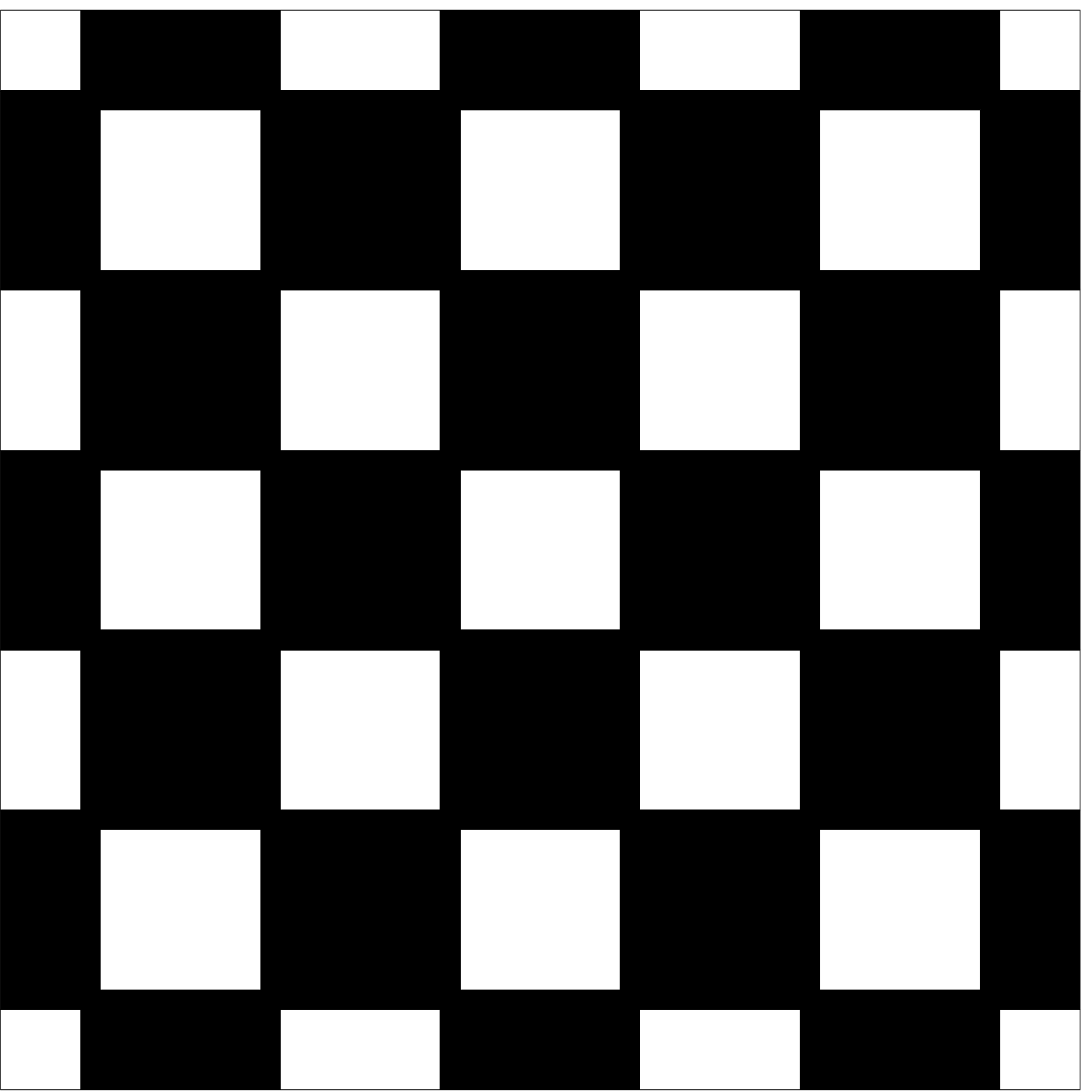}
  \includegraphics[width=35mm]{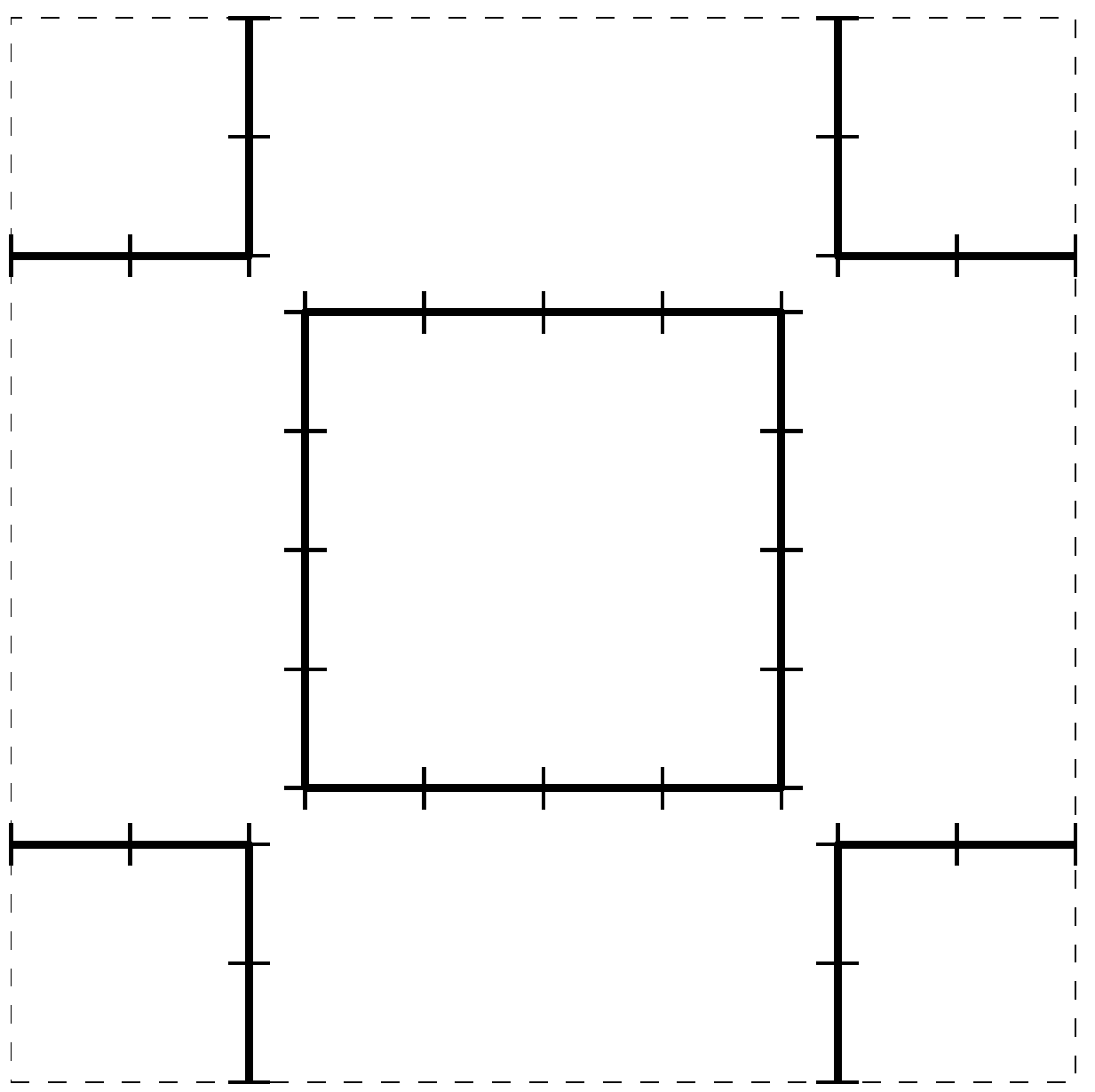}
\end{center}
\caption{Left: a cutout from ${\cal C}$ of a square array of
  square cylinders with area fraction $p_1=0.5$ and a unit cell ${\cal
    U}$ with a 16-panel coarse mesh on $\Gamma_0$. Right: the same
  thing for a staggered array of square cylinders, but with $p_1=0.4$
  and a 32-panel coarse mesh on $\Gamma_0$.}
\label{fig:boards}
\end{figure}

Laplace's equation is to be solved on a doubly periodic domain ${\cal C}$. The boundary conditions on the positively oriented interface
$\Gamma$ between the inclusion phase and the matrix phase are given in
Section~\ref{sec:over}. An average electric field $E_0$ of unit
strength is applied. The permittivity of the matrix phase is set to
$\epsilon_2=1$ so that the effective permittivity is equal to the
effective relative permittivity. From the repeated solution to this
problem for various $\epsilon_1=\sigma$ 
we obtain $\epsilon_{\rm eff}(\sigma)$. Two types of domains are investigated: the ``square
array of square cylinders'' and the ``staggered array of square
cylinders'' (a square array of cylinders with diamond-shaped
cross-sections, as studied for example in \cite{miltmcpedmcken,Hels00}), see Fig.~\ref{fig:boards}.

Boundary value problems on domains involving sharp corners may require
extreme resolution close to corner vertices, even when the demands for
overall accuracy are moderate. One has to be selective with the choice
of numerical method. We chose an integral equation based scheme. Such
schemes have the advantage that they can retain stability also in very
difficult situations.

Our particular choice of integral equation is standard -- a
single-layer equation~\cite{Gree06}. For its solution we use a novel
numerical method called {\it recursive compressed inverse
  preconditioning}. Conceptually this is a local multilevel technique
which makes a change of basis and expresses the non-smooth solution to
the single-layer equation in terms of a piecewise smooth transformed
layer density which can be cheaply resolved by polynomials.
Discretization leads to a block diagonal transformation matrix ${\bf
  R}$ (an inverse preconditioner) where the columns of a particular
block can be interpreted as special basis functions for the original
density in the vicinity of a corner vertex multiplied with suitable
quadrature weights. The blocks of ${\bf R}$ are constructed in a fast
recursion, $i=1,\ldots,n$, where step $i$ inverts and compresses
contributions to ${\bf R}$ involving the outermost quadrature panels
on level $i$ of a locally $n$-ply refined mesh. We emphasize that the
method is strictly numerical and fully automatic. There is no
separation of variables or eigenvalue analysis involved.

The recursive compressed inverse preconditioning method was originally
described in Ref.~\cite{Hels08} and further developed in
Refs.~\cite{Hels09a,Hels09b,Hels11a}. \ref{sec:AppA} below highlights
some of the method's features, relevant to the domain ${\cal C}$. A
fuller description will be included in a forthcoming
paper~\cite{Hels11b}.

\begin{figure}[t]
\begin{center}
  \includegraphics[height=60mm]{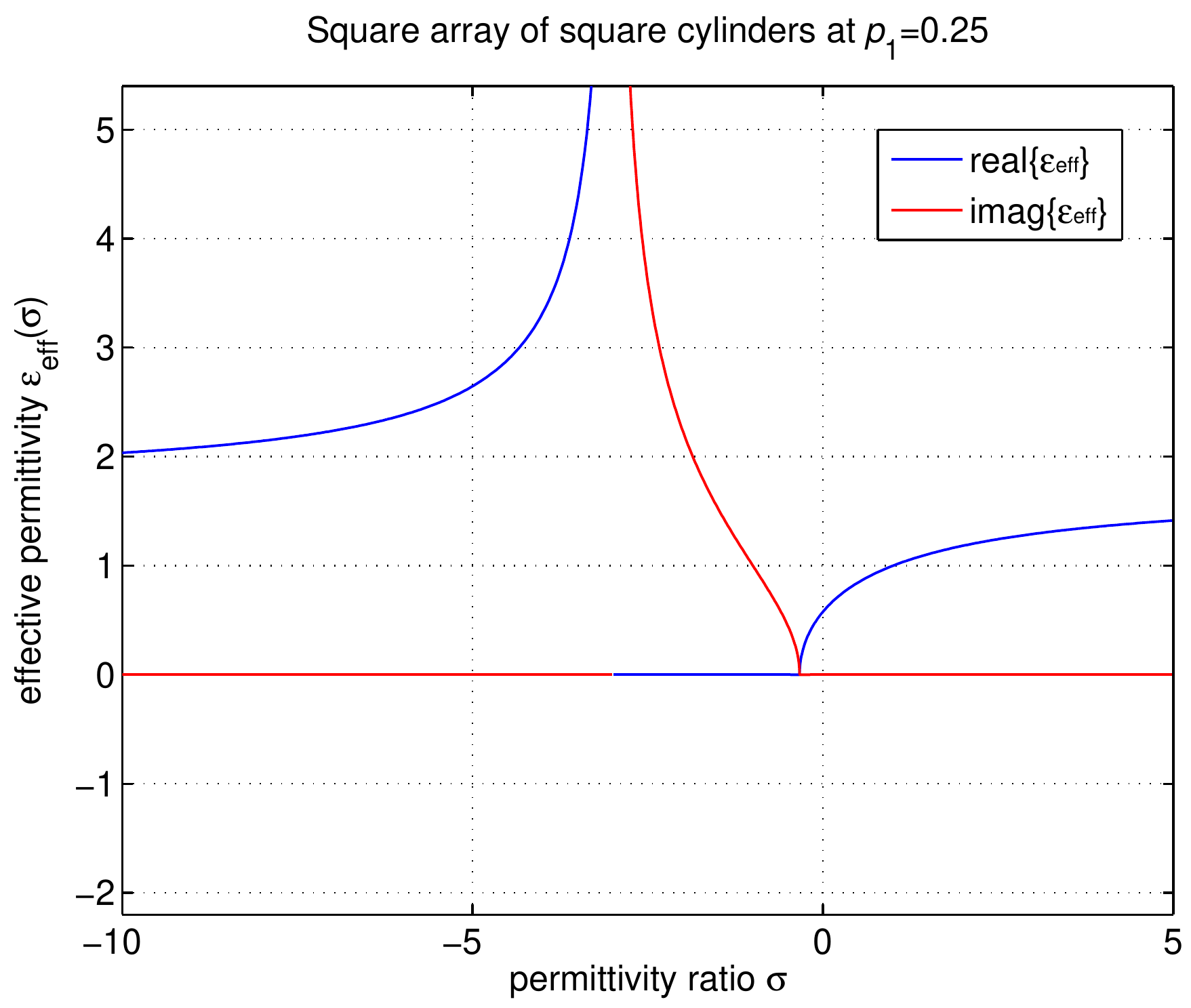}
  \hspace{5mm}
  \includegraphics[height=60mm]{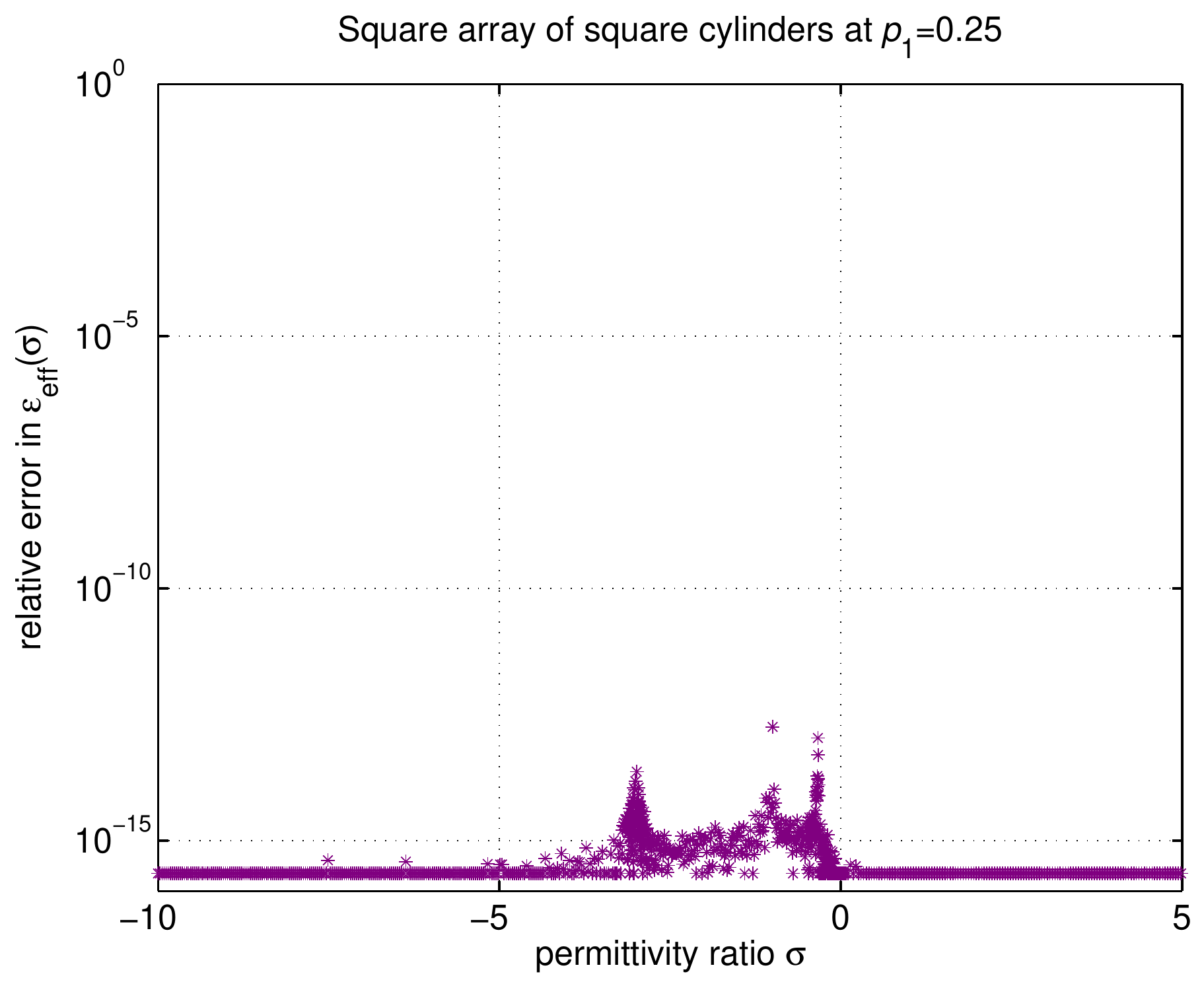}
\end{center}
\caption{Left: the effective relative permittivity of a square array of
  square cylinders at $p_1=0.25$. The curves are supported by 826
  adaptively spaced data points (not all values shown due to the
  setting of the axes). Right: the relative error with~(\ref{eq4}) as
  reference value.}
\label{fig:sqsq25}
\end{figure}

\subsection{Achievable accuracy}

We first compute $\epsilon_{\rm eff}(\sigma)$ for the square array of
squares at $p_1=0.25$ in the limit of $\sigma$ approaching the
negative real axis from the upper half-plane $\mathbb{H}$, as in the
example of Fig.~1. The exact result~(\ref{eq4}) is used as a benchmark. 
Fig.~\ref{fig:sqsq25} shows that the relative error is close to
machine epsilon (the upper bound due to rounding in floating point
arithmetic) except for in a neighbourhood of three points where it is
higher: the ends of the branch cuts at $\sigma=-3$ and $\sigma=-1/3$,
and at the singularity of the integral equation at $\sigma=-1$. 
This demonstrates that the problem of
computing $\epsilon_{\rm eff}(\sigma)$ for arrays of square cylinders
is well conditioned in general and that our scheme is stable.

\begin{figure}[t]
\begin{center}
  \includegraphics[height=60mm]{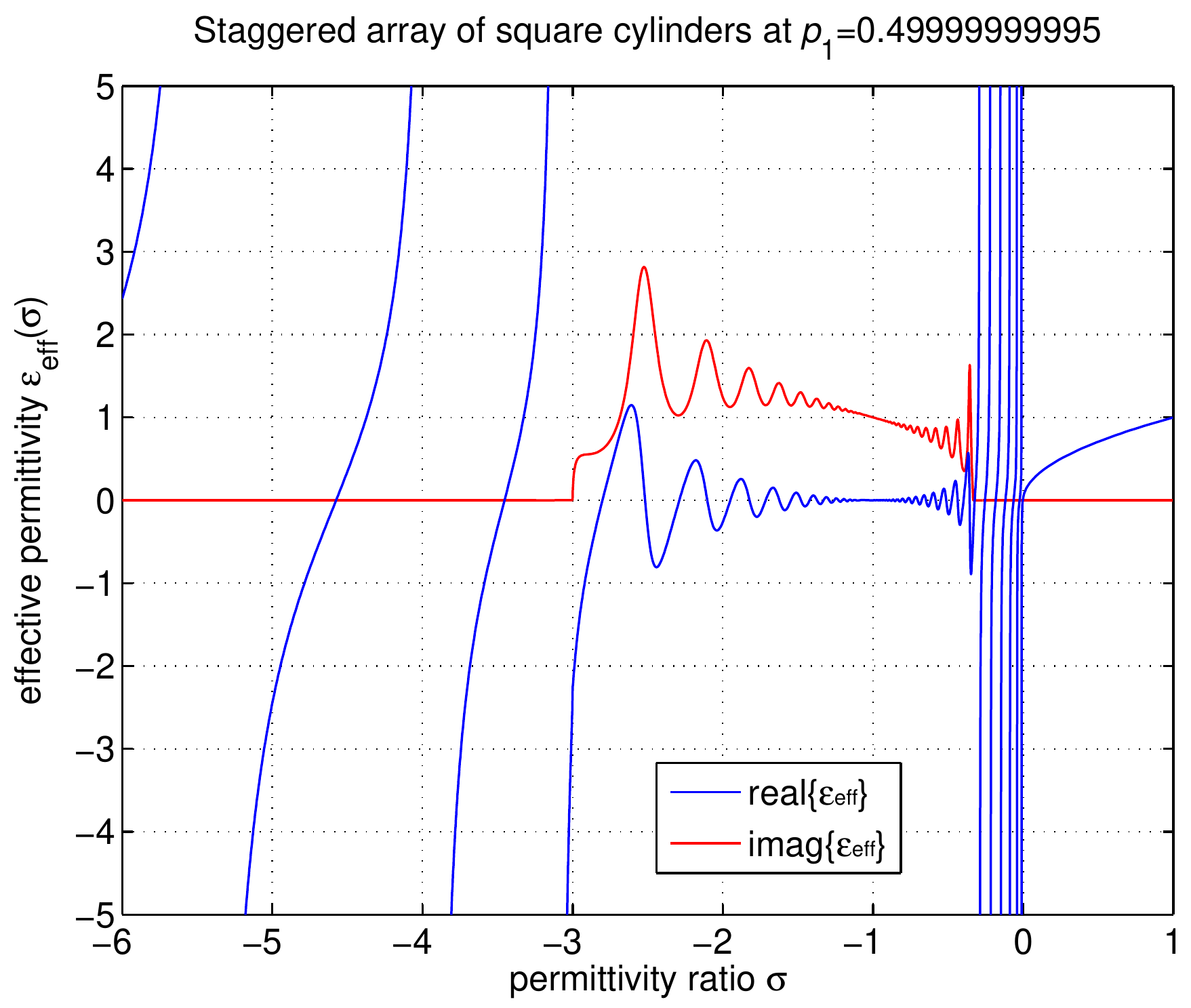}
  \hspace{5mm}
  \includegraphics[height=60mm]{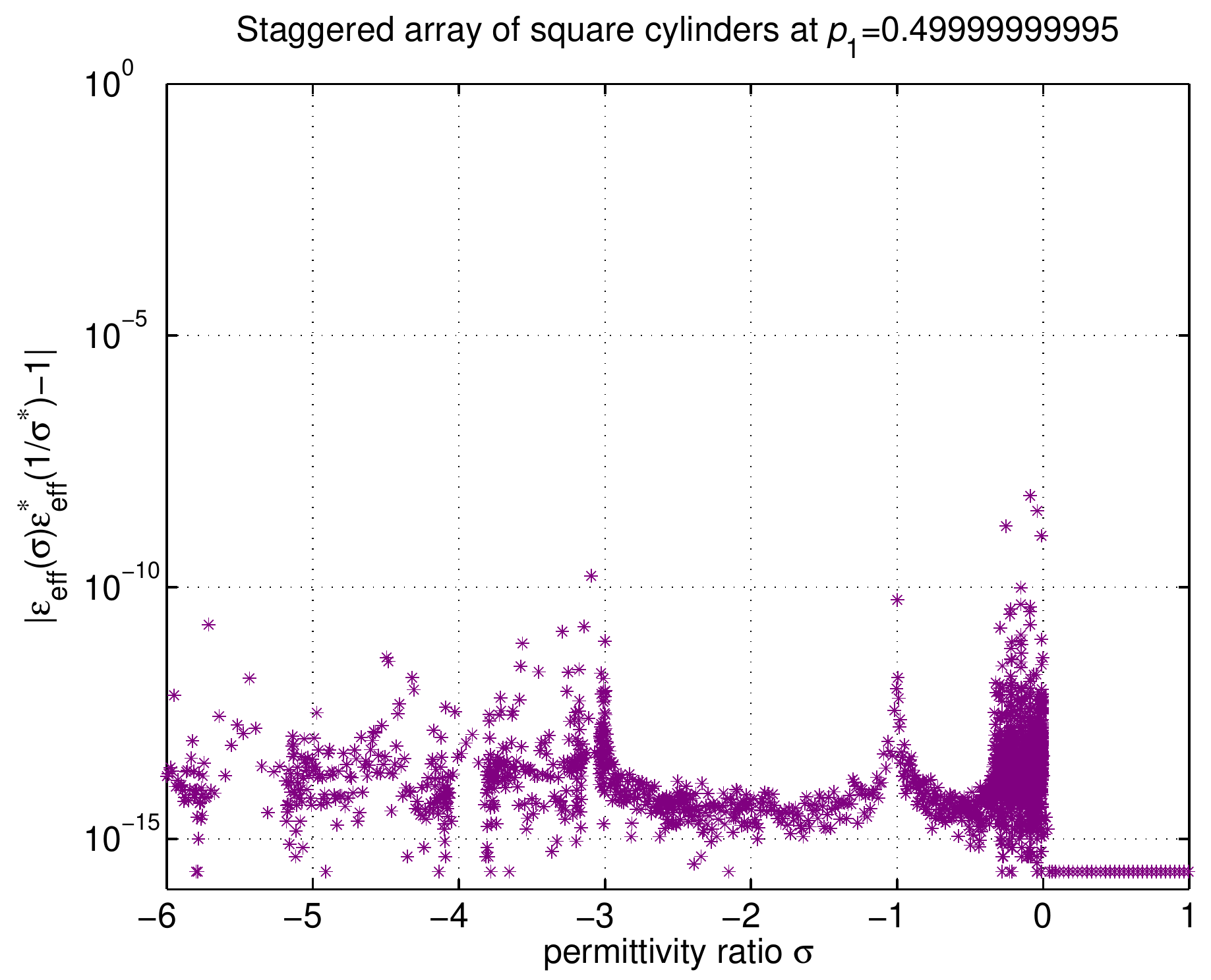}
\end{center}
\caption{Left: the effective relative permittivity of a staggered array of
  square cylinders at $p_1=0.49999999995$. The curves are supported by
  2006 adaptively spaced data points. Right: the absolute difference
  between the left- and the right hand side of~(\ref{eq2b}).}
\label{fig:stsqE}
\end{figure}

The staggered array of square cylinders at $p_1=0.49999999995$ is a
more challenging geometry than the square array of square cylinders at
$p_1=0.25$:
\begin{itemize}
\item there are more length scales involved,
\item $\epsilon_{\rm eff}(\sigma)$ varies more rapidly and has more poles
  and zeros,
\item there is no exact result to compare with.
\end{itemize}
The first problem is the least difficult. The multilevel property of
our numerical method should enable the resolution of almost
arbitrarily small separation distances between corner vertices. The
second problem is more severe. For $\epsilon_{\rm eff}(\sigma)$ close
to zero, one can expect cancellation in~(\ref{eq:eff1}) and the
relative accuracy should suffer. Furthermore, it is harder to resolve
wildly varying functions in floating point arithmetic than slowly
varying ones. The third problem is solved by using the extent to
which~(\ref{eq2}) is satisfied as an indicator of the relative error.
For this, since $\sigma$ and $1/\sigma$ lie on different sides of the
real axis and our numerical method takes limits from $\mathbb{H}$, we
use (\ref{eq2}) in the equivalent form
\begin{equation}
\epsilon_{\rm eff}(\sigma)\epsilon_{\rm eff}^{\ast}(1/\sigma^{\ast})=1\,,
\label{eq2b}
\end{equation}
where the `$\ast$' symbol to denotes complex conjugation.
Fig.~\ref{fig:stsqE} suggests that despite the difficulties,
typically, only a few digits are lost compared to the square array at
$p_1=0.25$. The numerics seem to give a relative precision of at least
$10^{-8}$ even for the most extreme values of $\epsilon_{\rm
  eff}(\sigma)$.

\begin{figure}[t]
\begin{center}
  \includegraphics[height=60mm]{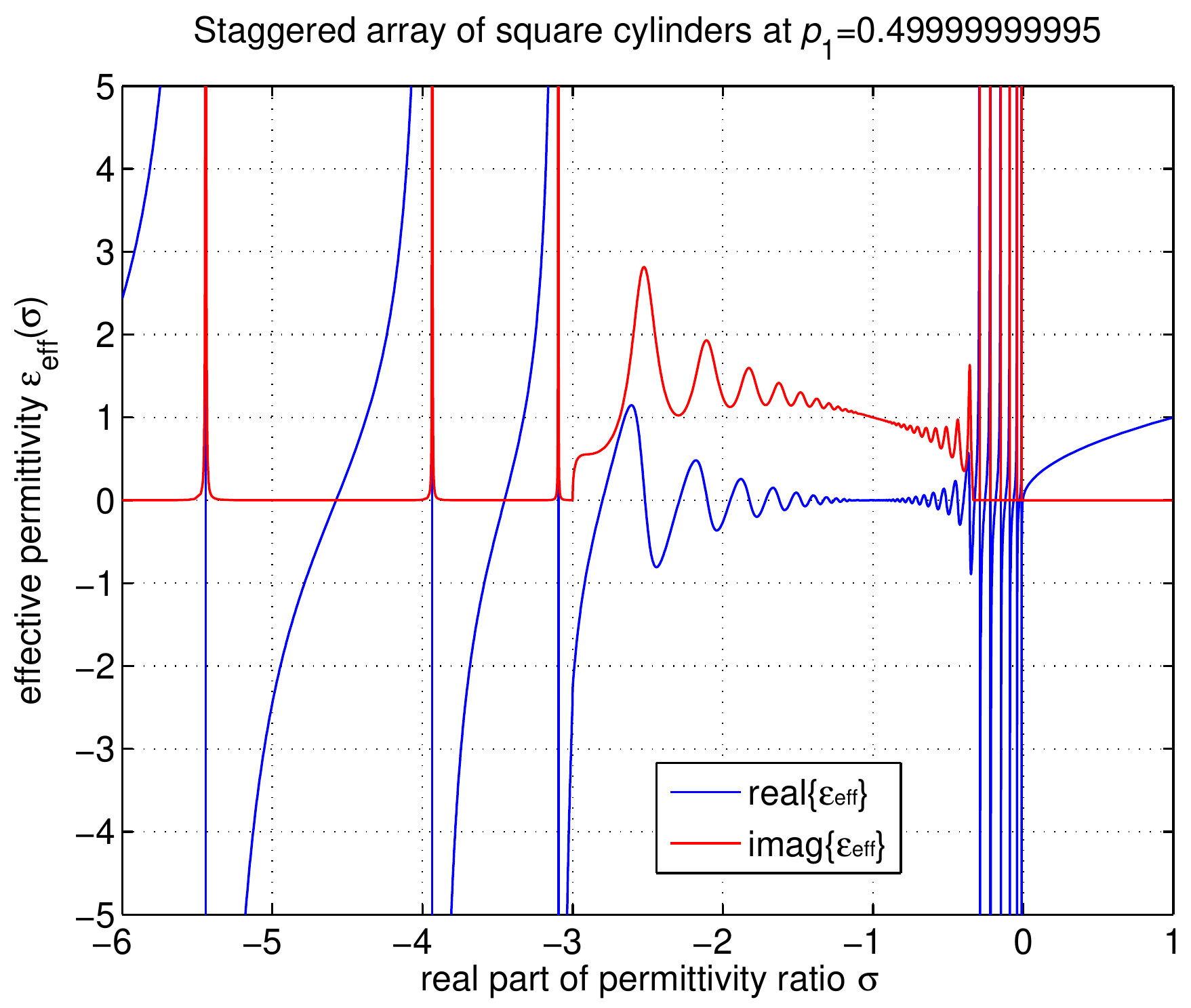}
  \hspace{5mm}
  \includegraphics[height=60mm]{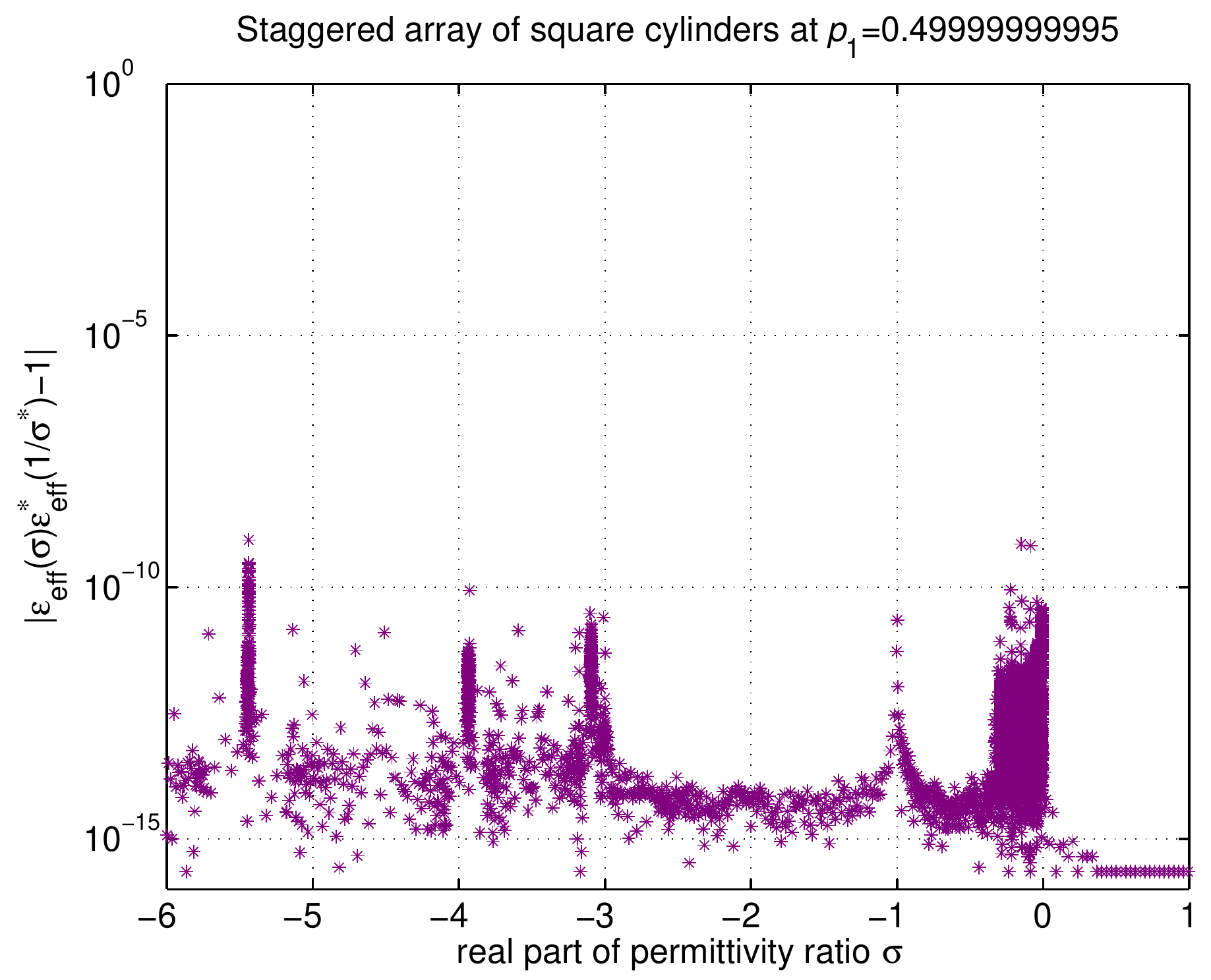}
\end{center}
\caption{Same as Fig.~\ref{fig:stsqE}, but $\sigma$ is 
  multiplied with a complex constant $1-{\rm i}\cdot10^{-5}$. The
  curves are supported by 3435 data points. The red curve is drawn on
  top of the blue curve.}
\label{fig:stsqI}
\end{figure}

For the interpretation of various limits it might also be of interest
to study $\epsilon_{\rm eff}(\sigma)$ for $\sigma$ some finite
distance into $\mathbb{H}$. Fig.~\ref{fig:stsqI} shows again the
staggered array of square cylinders at $p_1=0.49999999995$, but unlike
in Fig.~\ref{fig:stsqE} we have here interrupted the limit process at
$\sigma$ a relative distance of $10^{-5}$ away from the real axis.

\section{Animations of Spectral Evolution}
\label{sec:anim}

This section discusses the evolution of the variation of
$\epsilon_{\rm eff}(\sigma)$ with permittivity ratio $\sigma$ for the
square array of square cylinders and the staggered array of square
cylinders as the area fraction of squares, $p_1$, varies. The
discussion relates to two animations, called {\it Animation~1} and
{\it Animation~2}, which can be viewed at {\tt
  http://www.maths.lth.se/na/staff/helsing/animations.html}. To
facilitate viewing, each animation is available in four versions,
denoted {\tt A}, {\tt B}, {\tt C}, and {\tt D}. The versions have the
same content, but differ in image format and pixel resolution.

\begin{figure}
\begin{center}
\includegraphics[width=12cm]{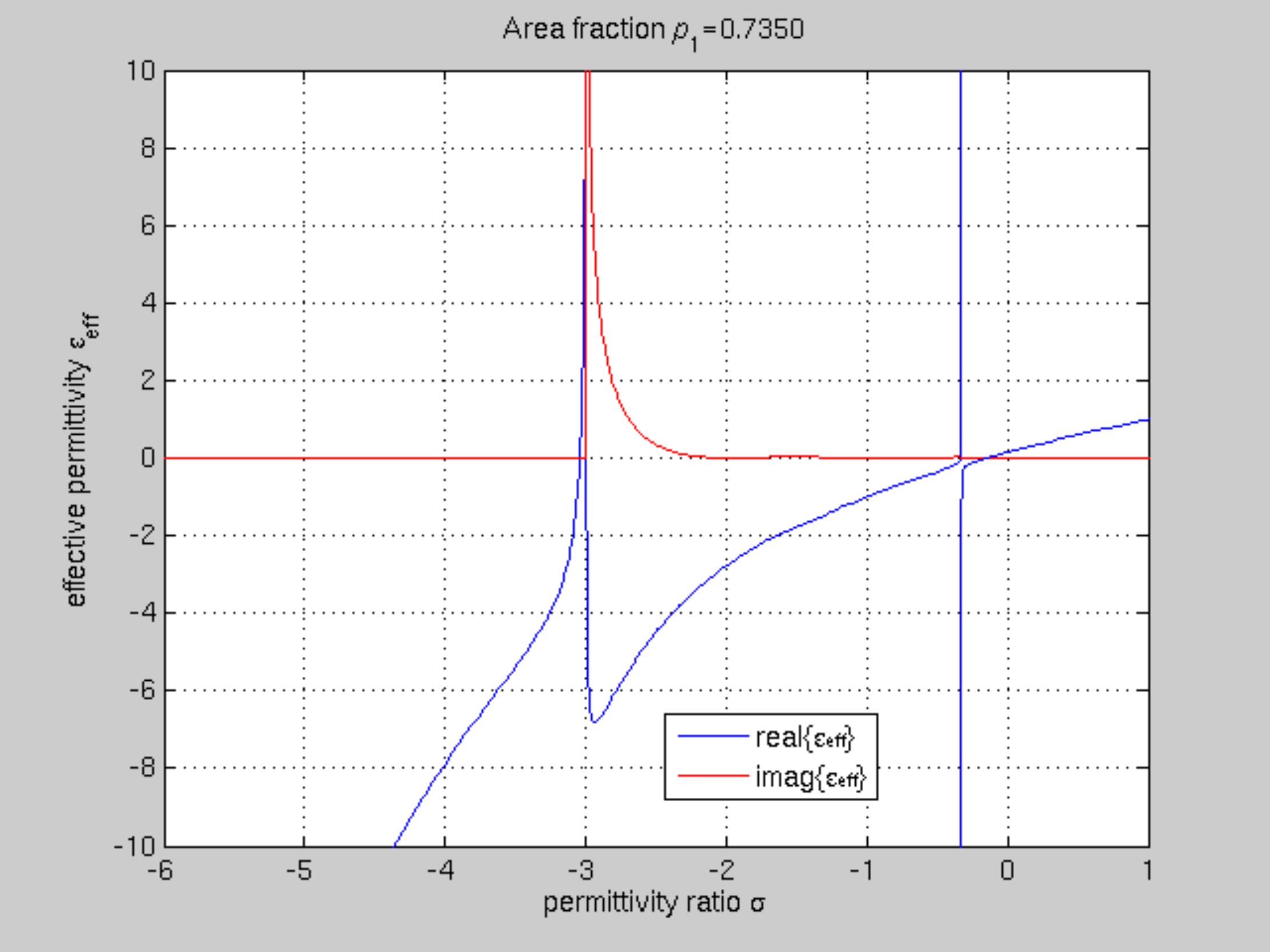}
\end{center}
\caption{Real (blue) and imaginary (red) parts of the effective
  relative dielectric permittivity for a square array of square prisms
  with area fraction $p_1=0.735$ as a function of permittivity ratio
  $\sigma$.}
\label{fig4-1}
\end{figure}

We first consider the evolution of the variation of $\epsilon_{\rm
  eff}(\sigma)$ as $p_1$ in the square array ranges from zero to
unity. This evolution is shown in Animation 1, from which a typical
frame is given in Fig.~\ref{fig4-1}. The most important feature of
Animation 1 is quite clear: for $\sigma$ real, and for all values of
$p_1$, and except at the poles, non-zero values of $\Im\{\epsilon_{\rm
  eff}(\sigma)\}$ are confined to the interval $-3\leq\sigma\leq-1/3$,
in accord with the suggestion of Hetherington and Thorpe
\cite{handth}. Of course, the value of this imaginary part is always
positive, if we restrict ourselves to composites without gain (for
which the imaginary part would be always negative).

Below the area fraction of 0.25, the real part of 
$\epsilon_{\rm  eff}(\sigma)$ is positive, and it develops its first pole
at this area fraction. It is interesting to
compare frames from Animation~1 and the left image of
Fig.~\ref{fig:sqsq25} with Fig.~\ref{fig1}; the results of the
mode-matching method clearly correspond to those of the new method,
but are capable of a resolution limited by the number of terms
employed in field expansions.

For $p_1>0.25$, the real part is negative between $\sigma=-3$ and
$\sigma=-1/3$, while the pole migrates to more negative values
of $\sigma$. For area fractions near $p_1=0.75$ (see
Fig.~\ref{fig4-1}), ``features'' which we call quasipoles develop from near $\sigma=-1$, and
one moves towards $\sigma=-3$, while the other moves towards
$\sigma=-1/3$. When they reach these values, and then are not muted by the absorbing nature
of the corners, they transform into
actual poles, which move towards $\sigma=-\infty$ and $\sigma=0$
respectively. At higher values of area fraction, more quasipoles
evolve from $\sigma=-1$ and give rise to additional actual poles when they
emerge from the branch-cut region.
$\Im\{\epsilon_{\rm eff}(\sigma)\}$ becomes small as $p_1\rightarrow
1$, while $\Re \{\epsilon_{\rm eff}(\sigma)\}$ tends towards $\sigma$,
apart from the increasingly numerous but increasingly narrow pole
regions.

For the staggered array of square prisms, Animation 2 illustrates the
behaviour of $\epsilon_{\rm eff}(\sigma)$ as a function of $\sigma$
for area fractions ranging from zero to 0.5 (with the behaviour for
$p_1$ in the range 0.5 to 1.0 following from that in the lower range
using Keller's Theorem~(\ref{eq2}). As has been commented in
Section~\ref{sec:over}, the interesting question is how the branch-cut
location (from $\sigma=-\infty$ to $\sigma=0$) for $p_1=1/2$ of
equation~(\ref{dyk}) can be reconciled with that (from $\sigma=-3$ to
$\sigma=-1/3$) predicted by Hetherington and Thorpe \cite{handth}
for $p_1$ arbitrarily near $1/2$. A
mechanism for this reconciliation was provided by one of the present
authors \cite{miltbook}: discrete sets of poles in
$-\infty\le\sigma\le-3$ and $\-1/3\le\sigma\le0$ were predicted to
become denser and denser as $p_1$ approached $1/2$, thus extending the
branch cut in the limit to that required by (\ref{dyk}). The accuracy
of this prediction is evident in Animation~2: poles develop from the
quasipoles generated at
$\sigma=-1$, and move left and right into the embryonic branch-cut
regions $-\infty\le\sigma\le-3$ and $\-1/3\le\sigma\le0$. The left
frame in Fig.~4 shows a stage in this evolution where $p_1$ is very 
close to $1/2$. Animation~2
makes the "nursery role" of the region around $\sigma=-1$
in the development of the spectrum much more evident (due
to larger amplitudes of the quasipoles) than does Animation~1.

The sensitivity of the spectral details for the staggered array near
the checkerboard configuration are very evident in Animation 2. As we
have commented in Section~\ref{sec:over}, the asymptotics of fields
near corners are the same in electromagnetism as in electrostatics
\cite{meixner}.  Thus, attempts such as that in \cite{pbeamsplitter}
to model the transition from electromagnetically reflecting structures
to electromagnetically transmitting structures as $p_1$ moves through
1/2 would require an adaptive and recursive method like that described
in Section~\ref{sec:numer} to be able to achieve sufficient accuracy.

\section{Discussion and Conclusions}
\label{sec:disc}

In this paper we have brought together rigorous mathematical results with numerical investigations of unprecedented accuracy. The latter have revealed the generality of the former, and have substantiated a conjecture of Hetherington and Thorpe \cite{handth} in a striking and conclusive way.

We conclude by commenting further on how the arguments and results we have presented can be implemented in a practical demonstration of morphological super-resolution, uniting the ideas of Kac  \cite{kac} and Pendry  \cite{pendry2000}. Such a demonstration would require the fabrication of a set of parallel cylinders with a square cross section (or  polygonal cross section). The cylinders do not have to be arranged in a geometrically-perfect array, and they do not have to be densely packed.  
They have to be made of a material which is essentially non-absorbing and with a negative permittivity or permeability.

These requirements suggest the set of cylinders be made of metal, of size 10 $\mu$m or larger, and be probed with wavelengths far greater than the cylinder size in the  far infrared or longer. Such cylinders are large by today's lithographic standards, and so it should be possible to accurately form their corners to achieve sub-wavelength accuracy. Going into the far infrared region diminishes metallic loss from its value in the visible and near-infrared \cite{bandw}. It is crucial that the metallic loss be very low, since the experimental signature we suggest be probed is enhanced absorption by a set of such cylinders over a wavelength interval in which the metal's permittivity ranges from say -1/3 to -3 (scaled relative to the permittivity of a host dielectric in which the cylinders are embedded). Note that the enhanced absorption of incident radiation detected will increase as more lines of cylinders are added to the set.

The experimental  result which would indicate morphological super-resolution is an enhanced absorption for wavelengths far in excess of the cylinder size, switching on and off at geometrically-determined limits described above, independent of the arrangement and area fraction of the cylinders. We stress however that such a demonstration would indicate the physical relevance of the ideas we have described for a  particular  system. The mathematical results we have described are of course rigorous, and the numerical examples of them we have given are highly accurate, so our demonstration of super-resolution for wavelengths arbitrarily larger than the size of the particles probed does not rely for its validity on experimental support. They may be applicable to governing equations other than the Helmholtz equation, for which the ideas of metamaterials and their applications are currently being explored \cite{seb}.

\ack Ross McPhedran acknowledges support from the Australian Research
Council's Discovery Projects and Centre of Excellence Schemes. Graeme
Milton acknowledges support from the National Science Foundation
through grant DMS-0707978. 

\appendix
\section{Key features of the numerical method}
\label{sec:AppA}

To keep the notation short we make no distinction between points or
vectors in a real plane $\mathbb{R}^2$ and points in a complex plane
$\mathbb{C}$. All points will be denoted $z$ or $\tau$.

\subsection*{The integral equation}

The potential function $V(z)$ in ${\cal C}$ is represented as a sum of
a driving term and a single-layer potential with density
$\rho(z)$~\cite{Gree94}. Enforcement of the boundary conditions on
$\Gamma$ leads to the Fredholm second kind integral equation
\begin{equation}
\rho(z)+\frac{\lambda}{\pi}\int_{\Gamma}\rho(\tau)
\Im\left\{\frac{n_zn^{\ast}_{\tau}\,{\rm d}\tau}{\tau-z}\right\}=
2\lambda\Re\left\{E^{\ast}_0n_z\right\}\,,\quad z\in\Gamma_0\,.
\label{eq:int1}
\end{equation}
Here $n_z$ is the outward unit normal of $\Gamma$ at $z$, $\Gamma_0$
denotes the restriction of $\Gamma$ to ${\cal U}$, and $\lambda$ is as
in~(\ref{eq:lmb}). Note that, as $\sigma\to-1$ we have
$\lambda\to\pm\infty$ and~(\ref{eq:int1}) is no longer a second kind
equation, but a first kind equation whose (unique) solvability is by
no means guaranteed. Therefore one can say that $\sigma=-1$ is a {\it
  singularity} of~(\ref{eq:int1}).

Once~(\ref{eq:int1}) is solved for $\rho(z)$ and under the assumption
that the inclusions do not overlap the unit cell boundary, the
effective relative permittivity in the direction of the applied
electric field can be computed from
\begin{equation}
\epsilon_{\rm eff}(\sigma)=1+
\int_{\Gamma_0}\rho(z)\Re\left\{E^{\ast}_0z\right\}\,{\rm d}|z|\,.
\label{eq:eff1}
\end{equation}

Depending on how the unit cell is chosen, the squares in the staggered
array may overlap the unit cell boundary. With the choice in
Fig.~\ref{fig:boards}, they certainly do. But since $\rho(z)$ is a
periodic function and identical on all squares one can circumvent this
problem by modifying~(\ref{eq:eff1}) so that it integrates $\rho(z)$
twice on the square at the center of the unit cell and ignores
$\rho(z)$ on the other squares.

\subsection*{Discretization}

We discretize~(\ref{eq:int1}) and~(\ref{eq:eff1}) using a Nystr{\"o}m
method based on composite 16-point polynomial interpolatory quadrature
and a parametrization $z(t)$ of $\Gamma$. The parameter $t$ is real.
See Ref.~\cite{Atki97} for a review of Nystr{\"o}m methods including
error analysis.

An initial coarse mesh that resolves the kernel of the integral
operator in~(\ref{eq:int1}) away from the corner vertices is
constructed on $\Gamma$, see Fig.~\ref{fig:boards}. The coarse mesh is
refined by subdividing those panels that neighbour corner vertices.
The subdivision is done $n$ times in a direction towards the vertices.
On quadrature panels which neighbour corner vertices we choose
quadrature nodes according to the zeros of certain Jacobi polynomials.
On remaining panels we choose quadrature nodes according to the zeros
of Legendre polynomials. Upon discretization on the refined
mesh~(\ref{eq:int1}) assumes the form
\begin{equation}
\left({\bf I}_{\rm fine}
+{\bf K}_{\rm fine}\right)\boldsymbol{\rho}_{\rm fine}
={\bf g}_{\rm fine}\,,
\label{eq:gen1}
\end{equation}
where ${\bf I}_{\rm fine}$ and ${\bf K}_{\rm fine}$ are square
matrices and $\boldsymbol{\rho}_{\rm fine}$ and ${\bf g}_{\rm fine}$
are column vectors. The vector ${\bf g}_{\rm fine}$ corresponds to the
discretization of the piecewise smooth right hand side.

Now the kernel $K(\tau,z)$ of the integral operator in~(\ref{eq:int1})
is split into two functions
\begin{equation}
K(\tau,z)=K^{\star}(\tau,z)+K^{\circ}(\tau,z)\,,
\label{eq:split}
\end{equation}
where $K^{\star}(\tau,z)$ takes care of corner interaction and
$K^{\circ}(\tau,z)$ can be viewed as the kernel of a compact integral
operator. The kernel split~(\ref{eq:split}) corresponds to an operator
split and the change of variables
\begin{equation}
\rho(z)=\left(I+K^{\star}\right)^{-1}\tilde{\rho}(z)
\end{equation}
makes~(\ref{eq:gen1}) read
\begin{equation}
\left({\bf I}_{\rm fine}+{\bf K}^{\circ}_{\rm fine}
\left({\bf I}_{\rm fine}+{\bf K}^{\star}_{\rm fine}\right)^{-1}\right)
\tilde{\boldsymbol{\rho}}_{\rm fine}={\bf g}_{\rm fine}\,.
\label{eq:gen2}
\end{equation}
This right-preconditioned equation corresponds to the discretization
of a Fredholm second kind equation with compact operators. The
solution $\tilde{\boldsymbol{\rho}}_{\rm fine}$ is the discretization
of a piecewise smooth function.

\subsection*{Compression}

The matrix ${\bf K}^{\circ}_{\rm fine}$, the density
$\tilde{\boldsymbol{\rho}}_{\rm fine}$, and the right hand side ${\bf
  g}_{\rm fine}$ in~(\ref{eq:gen2}) can be evaluated on the coarse
mesh without the loss of precision. Only $\left({\bf I}_{\rm
    fine}+{\bf K}^{\star}_{\rm fine}\right)^{-1}$ needs the refined
mesh for its accurate evaluation. This enables a compression
of~(\ref{eq:gen2}). We introduce the compressed weighted inverse
\begin{equation}
{\bf R}={\bf P}^T_W
\left({\bf I}_{\rm fine}+{\bf K}_{\rm fine}^{\star}\right)^{-1}{\bf P}\,.
\label{eq:R0}
\end{equation}
Here ${\bf P}$ is an unweighted prolongation operator that performs
panelwise 15th-degree polynomial interpolation in the parameter $t$ 
(which as we recall parameterizes $\Gamma$ through $z(t)$) from
points on the coarse mesh to points on the fine mesh when acting on
column vectors from the left. ${\bf P}_W$ is a weighted prolongation
operator. See Section~5 of Ref.~\cite{Hels09b}.

Substitution of~(\ref{eq:R0}) into~(\ref{eq:gen2}) and the use of some
relations between prolongation operators make~(\ref{eq:gen2}) assume
the form
\begin{equation}
\left({\bf I}_{\rm coarse}+{\bf K}_{\rm coarse}^{\circ}{\bf R}\right)
\tilde{\boldsymbol{\rho}}_{\rm coarse}={\bf g}_{\rm coarse}\,.
\label{eq:comp1}
\end{equation}
This equation, defined solely on the coarse mesh, will be used in our
computations.

\subsection*{Fast recursion for ${\bf R}$}

The construction of ${\bf R}$ from its definition~(\ref{eq:R0}) may be
a costly and unstable operation when the refined mesh has many panels.
The number of subdivisions $n$ needed to reach a given accuracy may
grow without bounds due to the singularities in $\rho(z)$ that arise
as $\sigma$ approaches certain values.

Fortunately, the construction of each block of ${\bf R}$, associated
with a corner of the square array or with a corner-meet of the
staggered array, can be greatly sped up and also stabilized via a
recursion. This recursion uses matrices ${\bf K}$ on local meshes
centered around corners or corner-meets. It would be going too far to
describe all the fine details of this procedure, but Sections~3.2
and~3.3 of Ref.~\cite{Hels11a} give a fairly good idea of how the
recursion is set up in the present context. A key step is the
(partial) conversion of the recursion into a non-linear matrix
equation. This equation is solved using a variant of Newton's method
relying on numerical homotopy to approach purely negative $\sigma$
from the upper half-plane $\mathbb{H}$. The ratio $\sigma$, which
enters into ${\bf K}$, is initially multiplied with a constant
$q=1-0.01{\rm i}$. The imaginary part of $q$ is reduced with a factor
of ten after each of the first 14 Newton iterations. Then $q$ is set
to unity and the iterations are continued until either a sharp
convergence criterion is met or a total of 30 iterations is reached. A
full description is given in Ref.~\cite{Hels11b}.

\end{document}